\documentclass[
reprint,
superscriptaddress,
amsmath,amssymb,
aps,
prl
]{revtex4-2}

\usepackage{bm}
\usepackage{graphicx}
\usepackage{hyperref}
\usepackage{mathrsfs}
\usepackage{xcolor}
\usepackage[T1]{fontenc}
\usepackage[utf8]{inputenc}

\hypersetup{
    colorlinks=true,
    linkcolor=blue,
    citecolor=blue,
    urlcolor=blue
}

\begin{document}

% Use the \preprint command to place your local institutional report
% number in the upper righthand corner of the title page in preprint mode.
% Multiple \preprint commands are allowed.
% Use the 'preprintnumbers' class option to override journal defaults
% to display numbers if necessary
%\preprint{}

%Title of paper
\title{When waves meet rays: Seismic vibrations and cosmic showers to test gravity}
% repeat the \author .. \affiliation  etc. as needed
% \email, \thanks, \homepage, \altaffiliation all apply to the current
% author. Explanatory text should go in the []'s, actual e-mail
% address or url should go in the {}'s for \email and \homepage.
% Please use the appropriate macro for each each type of information

% \affiliation command applies to all authors since the last
% \affiliation command. The \affiliation command should follow the
% other information
% \affiliation can be followed by \email, \homepage, \thanks as well.

\author{Aneta Wojnar}%\orcidlink{0000-0002-1545-1483}}
\thanks{Corresponding author}
\email[E-mail: ]{aneta.wojnar@uwr.edu.pl}
\affiliation{Institute of Theoretical Physics, University of Wroc\l aw, pl. Maxa Borna 9, 50-206 Wroc\l aw, Poland}

\begin{abstract}
We propose a novel laboratory test of gravity combining seismic-wave measurements with cosmic-ray muon detections. Quantum-gravity corrections to the anharmonic Debye model are derived, yielding a modified bulk modulus that encodes deviations from standard gravity. The usual dependence on density, a dominant source of uncertainty, is removed via muon tomography and seismic velocities measurement. We show that this setup can constrain gravity parameters at a level comparable to current laboratory experiments. Prospects for further improvements are briefly discussed.
\end{abstract}

\maketitle

A wide class of phenomenological approaches to quantum gravity, as well as classical gravity frameworks, predict corrections to microscopic physical quantities \cite{moussa2015effect,rashidi2016generalized,belfaqih2021white,mathew2021existence,hamil2021new,gregoris2022chadrasekhar,wojnar2023fermi,wojnar2024unveiling,Wojnar:2024hcg}. Incorporating the quantum structure of space-time typically leads to generalizations of the Heisenberg uncertainty principle, which may give rise to observable effects \cite{Pachol:2023bkv,Pachol:2023tqa,kozak2025refining}. In this context, the generalized uncertainty principle (GUP) has emerged as a powerful tool for capturing quantum-gravity-induced modifications \cite{kempf1995hilbert,maggiore1993generalized,maggiore1994quantum,chang2002exact,chang2002effect,ali2011minimal}. A common feature of such models is the appearance of a minimal length scale, typically associated with the Planck length, $L_P \sim \sqrt{\frac{\hbar G}{c^3}}$ \cite{bishop2020modified,bishop2022subtle,segreto2023extended}.

Within this framework, the GUP modifies the classical phase-space structure. In particular, the phase-space volume element is no longer trivial in order to remain invariant under time evolution. This deformation leads to a modified density of states \cite{Pachol:2024hiz}. As a result, fundamental statistical quantities are affected: the partition function must be redefined using the deformed measure, which in turn induces corrections to thermodynamic potentials, such as the Helmholtz free energy. These modifications propagate to observable quantities, altering the equation of state and other thermodynamic properties.

This feature opens the possibility of probing such effects in laboratory settings by analyzing the properties of materials, in particular solids. In what follows, we consider a Debye crystal at finite temperature, whose vibrational contribution to the Helmholtz free energy is given by
\begin{equation}\label{Fvib}
\mathcal{F}_\text{vib} =E_0+ 9pRT\left( \frac{T}{\theta_D}^3 \right)\int^\frac{\theta_D}{T}_0 \frac{x^2\mathrm{ln}(1-e^{-x})}{1+\bar\alpha(x)}dx,
\end{equation}
where $E_0= \frac{1}{2}\sum_{j=1}^{3pN} \hbar \omega_j$ is the zero-point vibrational energy within the harmonic approximation, $p$ denotes the number of modes, $\theta_D=\hbar\omega_D/k_B$ is the Debye temperature, and $x=\frac{\hbar\omega}{k_BT}$. Other symbols have their usual meaning. The total Helmholtz free energy is then given by
\begin{equation}
\mathcal{F}= E_\text{st} + \mathcal{F}_\text{vib},
\end{equation}
where $E_\text{st}$ is the static lattice energy at $T=0$.

On the other hand, $\bar\alpha(x)$ is, in general, an arbitrary function encoding corrections arising from a given model of gravity \cite{ali2009discreteness}. These corrections, following from the Liouville theorem, are incorporated into $\bar\alpha(x)$ and originate from a deformation of the elementary phase-space cell. This can be interpreted as a momentum-dependent modification of the unit cell associated with a particle if the correction is momentum-dependent. For generalizations, see \cite{Pachol:2024hiz}.

In the following, we restrict to the case where $\bar\alpha$ is a quadratic function of $\omega$, as this captures the most commonly studied modifications in the literature \footnote{That is, we focus on the quadratic momentum corrections to the Heisenberg Uncertainty Principle.}, giving rise to the quadratic GUP \cite{Bosso:2023aht}
\begin{equation}
   \Delta \mathcal{X} \Delta \mathcal{P} \geq \frac{\hbar}{2}(1+\beta_0\mathcal{P}^2),
\end{equation}
where $\Delta\mathcal{X}$ and $\Delta\mathcal{P}$ denote  position and momentum deviations, respectively, while $\beta_0$ is a gravity model parameter with the unit of inverse quadratic momentum.
Then, the zero vibrational energy $E_0$
arising from the harmonic approximation is also modified and takes the form \footnote{$\alpha\omega^2 = \beta_0\hbar^2/v_m^2 \mathcal{P}^2$, where $\beta_0$ is the quantum gravity parameter with the unit of inverse quadratic momentum, $v_m$ is the mean sound velocity, while $\mathcal{P}$ is the phonon momentum.}
\begin{equation}
    E_0 = \frac{9}{8}Rp\theta_D 
    \left( 1- \frac{2}{3}\alpha Np\left(\frac{2\theta_Dk_B}{\hbar}\right)^2 \right),
\end{equation}
where the last term is the effect of the deformed momentum phase. Similarly, one can distinguish the unmodified part of the Helmholtz vibrational energy,
\begin{equation}
\mathcal{F}_\text{vib}= R
pT\left[\frac{9}{8}\frac{\theta_D}{T}+
3\mathrm{ln}\left(w\right)-D\left(y\right)
\right],
\end{equation}
where $w = e^{-y}, \; y = \frac{\theta}{T}$, and
\begin{equation}
D\left(y\right) = 3 \left(\frac{T}{\theta_D}\right)^3\int^{\theta_D/T}_0 \frac{z^{3}dz}{e^z-1}
\end{equation}
is the Debye function, from the modified contribution, $F^\text{mod}_\text{vib}$, given by
{\small
\begin{align}
\mathcal{F}^\text{mod}_\text{vib}&=
-\frac{9 p R \alpha}{\theta^3}
\Bigg[
\frac{k_B^2 \theta^6}{12 \hbar^2}
+ \frac{k_B^2 T^2}{\hbar^2}
\Big(
\theta^4 \mathrm{Li}_2\left(w\right)
+ 4 \theta^3 T \mathrm{Li}_3\left(w\right)
\Big.\Big.\nonumber\\
&\Big.\Big.
+ 12 \theta^2 T^2 \mathrm{Li}_4\left(w\right)
+ 24 \theta T^3 \mathrm{Li}_5\left(w\right)
+ 24 T^4 \mathrm{Li}_6\left(w\right)
\Big)
\Bigg],
\end{align}}
where $
\mathrm{Li}_s(w) = \sum_{n=1}^{\infty} \frac{w^n}{n^s}$ are the polylogarithm functions.

Note that under the Debye's approximation, $\omega$ is a function of the volume $V$ providing that $\theta=\theta(V)$. Then, we can then easily find the pressure
\begin{align}
P&=-\left(\frac{\partial F}{\partial V}\right)_T=P_0
\\
&+\frac{pR\gamma}{V}
\left(\frac{9}{8}\theta+ 3\, T\, {D}\left(\frac{\theta}{T}\right)\right)\nonumber + \alpha P_\text{mod},
\end{align}
where $P_0$ is the pressure at $T=0$, $P_\text{mod}$ is a combination of the logarithmic and polylog functions of $\theta$ and $T$ while
\begin{equation}
  \gamma=  \gamma(V)= \frac{\partial\, \mathrm{ln} \theta}{\partial \,\mathrm{ln} V} 
\end{equation}
is the Gr\"uneisen parameter quantifying the relationship between the thermal and elastic properties of a solid. 

It is now straightforward to derive the isothermal bulk modulus
\begin{align}\label{kbulk}
K &= -V \left(\frac{\partial P}{\partial V}\right)_T  \\
  & = K_0+pR\frac{\gamma}{V} \Bigg[
     \left(1 - q + \gamma\right)
     \left(3\, T\, D\!\left(\frac{\theta}{T}\right)
     + \frac{9\, \theta}{8}\right) \nonumber \\
  &-12\, T\, D\!\left(\frac{\theta}{T}\right)\, \gamma \nonumber 
     + \frac{9\, \gamma\, \theta\Big(1-(\theta/T)^2\alpha \Big)}{e^{\theta/T}-1} + \alpha L_\text{mod}
  \Bigg],
\end{align}
where $ K_0$ is the bulk modulus at $T=0$, $q$ encondes the volume dependence of anharmonic effects in the lattice and it is given by $$q=\frac{\partial \mathrm{ln} \gamma}{\partial \mathrm{ln} V}, $$
while
\begin{equation}
    L_\text{mod}=\frac{9}{4}\left( \left(q-1- 3\gamma\right)\Big(L_\text{Li}-\frac{k_B^2\theta^3}{\hbar^2}\Big)
    + L_\text{Ln}
    \right)
\end{equation}
with
\begin{align}
L_\text{Li} = &-12\frac{k_B^2T^2\theta}{\hbar^2} \Bigg[
    \mathrm{Li}_2(w)
    + \frac{4}{y}\mathrm{Li}_3(w)
    + \frac{12}{y^2}\mathrm{Li}_4(w) \nonumber\\
    &+ \frac{24}{y^3}\mathrm{Li}_5(w)
    + \frac{24}{y^4}\mathrm{Li}_6(w)
\Bigg],\\
L_\text{Ln}=&\,4\frac{k_B^2T^2\theta}{\hbar^2} \big(q - 1 + \gamma\big)\,\ln\big(1 - w\big).
\end{align}

On the other hand, the bulk modulus can be obtained experimentally, as described in \cite{matsushima2024joint}. Let us consider a linear isotropic elastic material. Then, one can relate the material properties, such as bulk $K$ and shear $S$ moduli, together with its density $\rho$, to the $P$-wave velocity $v_P$ and $S$-wave velocity $v_S$, respectively:
\begin{align}
    v_P&=\sqrt{\frac{K+4 S/3}{\rho}}\label{vp},\\
    v_S&=\sqrt{\frac{S}{\rho}}\label{vs},
\end{align}
which can be rewritten as
\begin{equation}\label{Klab}
    K=\rho (v_P^2-\frac{4}{3}v_S^2).
\end{equation}
Although $P$- and $S$-wave velocities can be measured, the equations \eqref{vp} and \eqref{vs} alone are insufficient to uniquely determine $K$, $S$, and $\rho$, as they constitute a system of two equations with three unknowns. Knowledge of $\rho$ remains essential for extracting the remaining mechanical parameters. 

Most standard methods for determining material density rely, either directly or indirectly, on gravitational effects. Techniques based on weighing, such as geometric measurements or Archimedes' principle, explicitly depend on the gravitational acceleration, while even indirect approaches may involve quantities (e.g., pressure or elastic moduli) influenced by the gravitational environment. In particular, if pressure is used as a control or calibration variable, or if elastic properties (such as Young's modulus, bulk modulus, or shear modulus) enter the inference scheme, these quantities may themselves depend on the local effective gravitational field or on the underlying gravitational model through their coupling to stress distributions and equilibrium conditions. As a result, any gravity-induced modification of internal stresses or boundary conditions can propagate into the effective material response. Consequently, the inferred density may carry an implicit dependence on the underlying gravitational model, motivating the need for independent measurement strategies. 

One of such a strategy could be potentially muography. Muons, which are the relevant particles in this technique, have a mass of approximately $\sim10^{-28}$ kg, making them so light that gravitational effects on their propagation can, for the purposes of current applications, be safely neglected (see, however, the discussion below). As a result, muography is a method that is independent of gravity: it relies solely on electromagnetic muon-matter interactions, well described by the Bethe-Bloch equation \cite{bethe1953experimental}, rather than on measurements of gravitational forces.

In \cite{matsushima2024joint}, the bulk modulus \eqref{Klab} of aluminum was determined as $K=65.99$ GPa, using only the average density inferred from muon detection together with the $P$- and $S$-wave velocity measurements obtained from ultrasonic experiments on aluminum blocks \footnote{The errors were not provided apart from the density $\rho=2.58 \pm0.12$ g cm$^{-3}$. According to our analysis, $K = (65.8 \pm 3.1),\mathrm{GPa}$ solely from the density uncertainty.}. However, since the experiment was likely performed at ambient temperature $T=300$ K, we instead adopt the experimental reference value $K_0(V)=81.3$ GPa reported in \cite{gaudoin2002ab}.

Apart from this, we also calculated the Debye temperature for aluminium using the experimental data reported in \cite{matsushima2024joint}, in order to avoid relying on additional datasets obtained under different experimental conditions:
\begin{equation}
\theta_D = \frac{\hbar v_m}{k_B} (6\pi^2 n)^\frac{1}{3},
\end{equation}
where $n=M\rho/ N_A$, and $v_m$ denotes the mean sound velocity, determined from the measured $P$- and $S$-wave velocities as:
\begin{equation}
v_m=v_S\left( \frac{3}{2+\left(\frac{v_S}{v_P}\right)^3} \right)^\frac{1}{3}.
\end{equation}

The Debye temperature resulting from these experiments is $396\pm6$ K. It is worth noting that the Debye temperature is usually obtained experimentally from the specific heat curve at low temperatures \cite{kittel2018introduction}, which, however, may also depend on gravity, see e.g. \cite{riasat2023effect,Pachol:prep}. Another approach to determining the Debye temperature is temperature-dependent X-ray diffraction via the Debye-Waller factor \cite{horning1988debye}; however, this method relies on solving the Schr\"odinger equation for a quantum oscillator, which acquires quantum gravitational corrections \cite{Chang:2001kn}. This illustrates that muography, together with measurements of seismic vibrations, can be used to extract gravity-independent thermodynamic characteristics. We will use our value for $\theta_D$ to account for measurement uncertainties.

It is straightforward to determine the value of the parameter from \eqref{kbulk}. At ambient temperature, $T=300$K, and accounting for the experimental uncertainties, one obtains $\alpha=(1.438 \pm 2.94)\times 10^{-25}$ s$^2$, which corresponds to $\beta_0 \approx 1.53\times10^{50}$ s$^2$ kg$^{-2}$m$^{-2}$. The obtained value can be compared with results from other experiments, see, e.g., \cite{Bosso:2023aht}.

On the other hand, since in the temperature range $0-300$K no significant variations in \eqref{kbulk} are expected, we also consider, for instance, $T=10$K, for which we find $\alpha=(8.11 \pm 1.66)\times 10^{-24}$ s$^2$, corresponding to $\beta_0 \approx 8.64 \times 10^{50}$ s$^2$ kg$^{-2}$m$^{-2}$.

\begin{figure}[h]
     \includegraphics[scale=0.9]{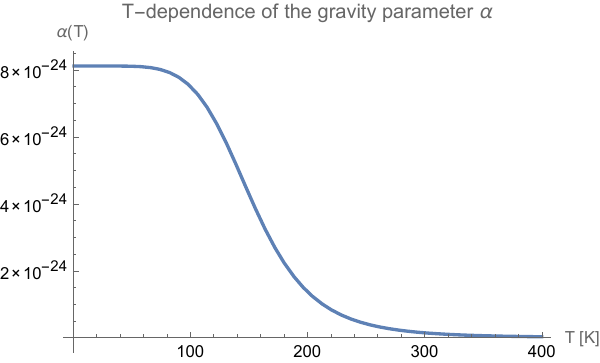}
     \caption{Expected values of the gravitational parameter $\alpha$ in low-temperature regime ($T<\theta_D$) for aluminum from the experiments performed in \cite{matsushima2024joint}.}
     \label{fig1}
\end{figure}

In Fig. \ref{fig1}, we illustrate this behavior, showing how the constrained value of the parameter $\alpha$ evolves with temperature. 
The sensitivity of the gravity parameter to the energy/temperature regime has already been discussed in \cite{kozak2021invariant,Lope-Oter:2023urz,Pachol:2023tqa,Pachol:prep}. The difference of nearly two orders of magnitude - partially also depending on the numerical precision - observed here (with a stronger bound at higher temperature), and potentially even larger at temperatures beyond those considered \footnote{Although the Debye model ceases to reliably describe crystal behaviour above the Debye temperature.} may admit a physical interpretation.

One possible explanation is that, at higher temperatures, as particles vibrate more rapidly, phonons may effectively acquire a mass, in the sense that correction terms appear in a momentum-dependent redefinition of the mass (see e.g. \cite{Visser:2005ss}). A similar phenomenon is known for photons (see e.g. \cite{fullekrug2004probing}). Alternatively, this behaviour can be interpreted in terms of a deformation of phase-space cells, which becomes more pronounced at higher momenta. This effect may be also reformulated in terms of a modified dispersion relation \cite{hossenfelder2006note}. Such modifications are often motivated by attempts to describe quantum properties of spacetime \cite{amelino2013quantum,addazi2022quantum}, e.g. through noncommutativity \cite{Borowiec:2010yw,Aschieri:2020yft,Aschieri:2017ost}. A key phenomenological implication concerns the fate of relativistic symmetries \cite{amelino1998tests}, leading in particular to scenarios of Lorentz Invariance Violation (LIV), in which modified dispersion relations are not preserved under standard Lorentz transformations. 

The impact of LIV on cosmic showers has been investigated in \cite{trimarelli2022constraining}. It was shown that LIV affects the number of muons produced in the atmosphere, while leaving their energy essentially unchanged, which could indirectly influence the inferred densities of the aluminum block. Therefore, muography can still be regarded as a gravity-independent method for determining densities and other elastic properties of materials.

Nevertheless, in order to fully exploit the potential of the proposed method, the experiment described in \cite{matsushima2024joint} requires further development and refinement. First, the setup should be improved by significantly increasing the number of detection channels (only 3 are used in \cite{matsushima2024joint}), thereby enabling a much higher measurement resolution. In addition, implementing a multilayer detector with signal readout from both ends of each segment would allow for precise reconstruction of the particle trajectory, leading to a more accurate determination of its direction as well as improved timing measurements \cite{bielewicz2021conceptual,bielewicz2023practical,bielewicz2018mcord}. Altogether, these upgrades would enhance the determination of density and elastic moduli.

Moreover, it is important to account for previously unreported uncertainties in the seismic velocity measurements which we use here, which directly affect the inferred thermodynamic properties of the material. Another relevant improvement would be to perform the experiment under controlled environmental conditions, for instance by integrating the cosmic-ray detector setup with a climatic chamber, allowing measurements to be systematically repeated at different temperatures. This is particularly relevant, as observing an effect such as the one shown in Fig. \ref{fig1} could potentially indicate sensitivity to quantum gravity effects.

\section*{Acknowledgements}
\textit{ For those who wander the thin line between dream and abyss: may courage guide us to a quiet shore.}

AW thanks Juan \'Angel Sans Tresserras and Marcin Bielewicz for interesting discussions. 

This article is based upon work from COST Action FuSe, CA24101, supported by COST (European Cooperation in Science and Technology).
% \\[2ex]

%  lub 

% For the dreamers who walk the edge of the abyss - may their courage lead them to peace.

\bibliographystyle{apsrev4-1}
\bibliography{biblio}

\end{document}